\documentstyle[prd,aps,preprint,eqsecnum,tighten]{revtex}

\def\frac#1,#2{{#1\over #2}}

\def\fr#1,#2{{#1\over #2}}
\def\lan{\langle}
\def\ran{\rangle}
\def\bra#1{\lan#1|}
\def\ket#1{|#1\ran}
\def\ha{{1\over 2}}

\begin{document}

\preprint{\vbox{\hfill SMUHEP/03--01}}


\title{Calculations in the Light-Cone Representation\footnote{\uppercase{T}his work is supported by the \uppercase{U.S.} \uppercase{D}o\uppercase{E}}}

\author{Gary McCartor}

\address{Department of Physics, \\ 
SMU, \\
Dallas, Texas 75275\\
E-mail: mccartor@mail.physics.smu.edu}  


\maketitle

\begin{abstract}
For the problem of calculating bound states in quantum field theory, the light-cone representation offers advantages over the more common equal-time representation.  It also has subtleties and disadvantages compared to the equal-time representation.  If current efforts to use the light-cone representation to solve for the properties of hadrons in QCD are to succeed, at least two problems have to be solved: we must find the induced operators; we must develop an effective procedure of regularization and renormalization.  In this paper I will try to explain what an induced operator is and say what we know about them and will report on recent attempts to develop an effective procedure of regularization and renormalization.
\end{abstract}


\section{Induced Operators}

The procedure of light-cone quantization is to specify initial conditions for the fields (the canonical commutation relations) on the surface $x^+ = 0$ then solve for the fields as a function of $x^+$~\cite{dbp}.  For the bound state problem this is done by solving
$$
(P^+ P^- + P_\perp^2) \ket{\psi} = M^2 \ket{\psi}
$$
In light of the complications I will discuss below it is appropriate to provide some justification for thinking that light-cone quantization can have advantages.  Perhaps the three most often quoted advantages are: boost invariance (boosts in the z-direction are kinematical in the light-cone representation); a much less complicated vacuum; and much simpler eigenstates.  These last two properties are related and, from the point of view of this paper, are the important advantages.  In addition to these advantages there are problems which either do not occur in the equal-time representation or are different in form.  In this section I shall discuss the problem of induced operators; in the next section I shall discuss the problem of regularization and renormalization.

I shall first discuss the case of the Schwinger model; the one case of an induced operator which is understood in complete detail.  In light-cone gauge, the operator solution to the Schwinger model is given by~\cite{nm}
$$
     \Psi_+ = Z_+ e^{\Lambda_+^{(-)}}\sigma_+ e^{\Lambda_+^{(+)}}
$$
$$
      \Lambda_+ = -i2\sqrt{\pi}({\eta}(x^+) + \tilde{\Sigma}(x^+,x^-))
$$
$$
       Z_+^2 = \fr{m^2e^\gamma},{8\pi\kappa}
$$
$$
     \Psi_- = Z_-e^{\Lambda_-^{(-)}}\sigma_- e^{\Lambda_-^{(+)}}
$$
$$
        Z_-^2 = \fr{\kappa e^\gamma},{2\pi}
$$
$$
      \Lambda_- = -i2\sqrt{\pi}\phi(x^+)
$$
$$
        A_+ = \fr{2},{m} \partial_+ ({\eta} + \tilde{\Sigma})
$$
Here, $\tilde{\Sigma}$ is a massive pseudoscalar (free) field, the physical field which creates the Schwinger particle.  $\phi$ is an auxiliary positive metric field while $\eta$ is a negative metric auxiliary field.  This solution will be found by quantizing at equal-time, on $x^- = 0$ or on $x^+ = 0$.  If we quantize at $t = 0$ (or on $x^- = 0$) no special care is needed to include $\phi$ and $\eta$ in the solution; they will automatically be included by using standard techniques.  It is clear, however, that if  we are to include them in the solution when quantizing on $x^+ = 0$, which we must do if we are to get the correct answer, some care must be taken.

Before discussing how the $x^-$-independent fields appear in the light-cone representation, let me briefly discuss what their role is in the solution --- that is, what bad things will happen if we leave them out.  $\eta$ changes the singularity in the $\Psi_+$ two-point function from an unacceptable $e^{-2\gamma}\fr{4},{m^2} \fr{1},{\epsilon^+\epsilon^-}$, if we keep only the $\tilde{\Sigma}$ field, to the correct $\fr{1},{2\pi\epsilon^-}$.  $\phi$ and $\eta$ are the only fields in the problem which carry a charge, so leaving them out would be doing electrodynamics without charges.  Finally, $\phi$ and $\eta$ provide the only operators that create states which can dress the bare vacuum.  As is generally known, the Schwinger model has a one-parameter vacuum ambiguity --- the $\theta$-ambiguity.  All of the $\theta$-states are dressed by operators from the $\phi$ and $\eta$ fields, but in the light-cone representation they are still far simpler than the corresponding vacua in the equal-time representation where they are more heavily dressed by $\phi$ and $\eta$ and are also dressed by $\tilde{\Sigma}$.  If we leave out $\eta$ or $\phi$ we will not be able to form a $\theta$-state and will therefore not be able to implement invariance under the large gauge transformations.  If we include $\phi$ and $\eta$ we can form a correct vacuum, and, for instance, can easily calculate the expected chiral condensate:
$$
 \langle\Omega (\theta )|\bar{\Psi}\Psi|\Omega (\theta )\rangle = -{\frac{m},{2\pi}}
{\rm e}^\gamma \cos\theta
$$

The fields $\phi$ and $\eta$ appear in light-cone quantization as integration constants.  Light-cone quantization typically involves the solution of differential constraint relations and we must be careful with the boundary conditions.  For the Schwinger model these constraint relations are
$$
    2 \partial_-^2 A^- = - J^+
$$
$$
   i \partial_- \Psi_- = 0
$$
Looking at the solution given above we see that $\phi(x^+)$ is one of the integration ``constants" associated with solving the constraint equation for $A^-$ (the other integration ``constant" is zero), while the entire field, $\Psi_-$, is the integration constant associated with solving the constraint equation for $\Psi_-$.  There must be some principal which determines the value of these integration constants.  It is that these fields should be canonical at space-like separations.  It is best to think of it as follows: there is only one operator solution although it may be written in an infinite number of bases; if we quantize at equal-time we know we will have canonical fields at space-like separations; changing the basis does not change the algebra.  The thing that makes space-like separations special is that each point is causally disconnected from every other point whereas at light-like separations the algebra can become ill defined as we see by looking at the full solution to the Schwinger model given above.  So far we have found that the integration constants are not generally zero but we do not yet have an induced operator.  That is because the integration constants do not change the spectrum of the Schwinger model.  They will effect the spectrum and we will find an induced operator if we add a bare mass to our Lagrangian and consider the massive Schwinger model.

With a nonzero bare mass, $\mu$, the constraint relation for $\Psi_-$ becomes
$$
\partial_- \Psi_- + i\ha\mu\Psi_+ =0 
$$
This we solve as~\cite{m}
$$
   \Psi_- = \Psi_-^0(x^+) -\int i\ha\mu\Psi_+ dx^-
$$
$\Psi_-^0(x^+)$ is again determined by the requirement that $\Psi_-$ be canonical at space-like separations:
$$
 \{\Psi_-(x),\Psi_-(x + \epsilon_{SPACE})\} = \delta(\epsilon_{SPACE})
$$
and the physical subspace requirement (I am glossing over the point that while the requirement of canonical fields at space-like separations determines the integration constants, that requirement may not be particularly convenient to use in practical cases).  Now, the operator, $\bar{\Psi}\Psi$ will include a cross term involving the integration constant which will act in the physical subspace:
$$
   \bar{\Psi}\Psi \supset (\Psi^{0*}_- \Psi_+ + \Psi^*_+ \Psi^0_- )
$$ 
This term leads to the linear growth of the mass squared of the Schwinger particle with the bare mass~\cite{m}:
$$
         \bra{p} P^+ \delta P^- \ket{p} = - 4 \pi\mu \bra{\Omega}\bar{\Psi}\Psi\ket{\Omega} = 2 m \mu e^\gamma \cos{\theta}
$$
The part of $\bar{\Psi}\Psi$ which depends on the integration constant is an example of an induced operator.  It is very important to understand that it is not a new operator: if we quantized at equal-time it would automatically be included in our Hamiltonian.  The only reason for giving it the special name of an induced operator is that when we quantize on the light-cone, correctly including it in the dynamics requires that we take special care with the integration constants which are associated with the constraint relations --- the operator, which acts in the physical subspace, is induced by the integration constants even though those are auxiliary fields.

There will be such operators for many realistic field theories including QCD.  They have not all been worked out.  For speculation as to the one in QCD which most closely corresponds to the one in the Schwinger model see ref. [4].  There will be other induced operators in QCD in addition to that one.

\section{Regularization and Renormalization}

For several years Brodsky, Hiller and I have been trying to develop procedures for performing nonperturbative renormalization~\cite{bhm}; more recently we have also been collaborating with Franke, Prokhvatilov and Paston on the same topic.  The Basic idea is to add Pauli-Villars fields to regulate the theory in a way which preserves as many of the sacred symmetries as we can; were we must break these symmetries we must add counter terms.  Only after the theory is finite do we truncate the Fock space so as to get a problem we can solve.  Since there is a finite target which we hope to approximate, the validity of this last step is a question of accuracy rather than symmetry.  If our approximate answer lies sufficiently close to the answer which preserves the symmetries it does not matter if the small difference breaks the symmetries.

Most of our studies so far have been on Yukawa-like theories so that we have no infrared problem to face and we do not have to worry about protecting gauge symmetries; we are now extending our calculations to QED so some of those complications will now have to be faced.  We have learned two important lessons from the studies so far performed.  The first lesson is that at some point there is always a rapid drop off of the projection of the wave function onto higher Fock sectors.  Just where this occurs depends on the theory, the coupling constant and the value of the Pauli-Villars masses.  At weak coupling only the lowest Fock sectors are significantly populated.  At stronger coupling more Fock sectors will be populated but eventually the projection onto higher sectors will fall rapidly.  The projection onto the higher Fock sectors also grows as the values of the Pauli-Villars masses increase.  The rapid drop off in the projection of the wave function onto sufficiently high Fock sectors is the most important reason why we do our calculations in the light-cone representation.  For any practical calculations on realistic theories we have to truncate the space and we must have a framework in which that procedure can lead to a useful calculation.  The rapid drop off in the projection of the wave function will not happen in the equal-time representation mostly due to the complexity of the vacuum in that representation.  These features can be explicitly demonstrated by setting the Pauli-Villars masses equal to the physical masses.  In that case the theory becomes exactly solvable~\cite{bhm2}.  The spectrum is the free spectrum and the theory is not useful for describing real physical processes due to the strong presence of the negative normed states in physical wave functions but it still illustrates the points we have been trying to make.  In that case the physical vacuum is the bare light-cone vacuum while it is a very complicated state in the equal-time representation.  Physical wave functions project onto a finite number of Fock sectors in the light-cone representation but onto an infinite number of sectors in the equal-time representation.  While the operators that create the physical eigenstates from the vacuum are more complicated in the equal-time representation than in the light-cone representation the major source of the enormous complication of the equal-time wave functions is the equal-time vacuum.  As the Pauli-Villars masses become larger than the physical masses, the light-cone wave functions project on to more of the representation space and more so as the coupling constant is larger and the Pauli-Villars masses are larger, but the wave functions remain much simpler than in the equal time representation and to the extent we can do the calculations there is always a point of rapid drop off of the projection onto higher Fock sectors.

The other lesson that we have learned is not really a new lesson: we should not plan to take the masses of the Pauli-Villars fields all the way to infinity even if we have the computational ability to do so.  That restriction comes from the problem of uncancelled divergences which will always occur when we do a nonperturbative calculation in a truncated representation space.  To illustrate the problem, consider the nonperturbative calculation of an anomalous magnetic moment of a fermion dressed with a single boson.  The result has the form
$$ 
\mu^\prime = {g^2 [finite quantity] \over 1 + g^2[finite quantity] + g^2[finite quantity] \log \mu_2} 
$$
where $\mu_2$ is the Pauli-Villars mass scale.  If we let $\mu_2$ go to infinity we will get zero.  That would not happen in perturbation theory: since the numerator is already order $g^2$ we would use only the 1 from the denominator and might get a nonzero result.  In order $g^4$ we would use the divergent term in the denominator but there would be new terms in the numerator which would contain canceling divergences.  This is the problem of uncanceled divergences. The solution is to keep the Pauli-Villars masses finite.  We think of it this way:  If the limit of infinite Pauli-Villars masses would give a useful answer then there must be some finite value which would also give a useful answer.  The question is whether we can use a sufficiently large value.  To answer that question we must consider that there are two types of error associated with the value of the Pauli-Villars masses.  The first type of error results in having these masses too small; then our wave function will contain too much of the negative normed states.  That type of error goes like
$$
    E_1 \sim {M_1/M_2}
$$
Where $M_1$ is the mass of the heaviest physical particle and $M_2$ is the mass of the lightest Pauli-Villars particle.  The other type of error results from having the Pauli-Villars masses too large, in which case our wave function will project significantly onto the parts of the representation space excluded by the truncation.  That error can be roughly estimated as
$$
    E_2 \sim {\langle \Phi_{+}^\prime|\Phi_{+}^\prime\rangle \over \langle \Phi_{+}|\Phi_{+}\rangle}
$$
where $|\Phi_{+ }^\prime\rangle$ is the projection of the wave function onto the excluded sectors.  In practice this quantity can be estimated by doing a perturbative calculation using the projection onto the first excluded Fock sector as the perturbation.  If both types of error are small, we can do a useful calculation; otherwise not.  But, due to the eventual rapid fall off of the projection of the wave function onto higher Fock sectors, we believe it will always be possible, in principle, to include enough of the space to do a useful calculation.  Computational limitations might mean that it would not be possible in practice.

We believe that the calculations have progressed to the point where we need to attempt a calculation for a problem to which we know the answer.  We are therefore attempting a nonperturbative calculation of the electron's magnetic moment.  To perform this calculation successfully we must overcome three problems: the problem of uncanceled divergences, which I have just been discussing; the problem of the appearance of new divergences which do not occur in perturbation theory; and the problem of maintaining gauge invariance (not a trivial problem).  We believe that we have techniques to overcome each of these problems and we hope to report a successful calculation in the near future.

\end{document}